\documentclass[11pt,a4paper]{article}

\usepackage{graphicx}
\usepackage{amsmath}
\usepackage{amssymb}
\usepackage{url}
\usepackage{microtype}
\usepackage{setspace}
\usepackage{caption}
\usepackage[hidelinks]{hyperref}
\usepackage{wrapfig}
\usepackage{titlesec}
\usepackage{cite}
\usepackage[top=0.95in, bottom=0.9in, left=1.1in, right=1.1in]{geometry}
\usepackage[utf8]{inputenc}
\usepackage{xcolor}
\usepackage{MnSymbol,wasysym}
\usepackage{listings}
\usepackage[normalem]{ulem}

\hypersetup{
   pdfauthor = {Ga\"el Richard, Vincent Lostanlen, Yi-Hsuan Yang, Meinard M\"uller},
   pdftitle = {Model-Based Deep Learning for Music Information Research}
}

\def\maintextlinespacing{1.59}
\def\referenceslinespacing{1.46}
\def\appendixlinespacing{1.59}

\titlespacing*{\section}
{0pt}{1.2ex}{0.9ex}
\titlespacing*{\subsection}
{0pt}{1.2ex}{0.9ex}

\newenvironment{itemizePacked}{
\begin{itemize}
  \setlength{\itemsep}{1pt}
  \setlength{\parskip}{0pt}
  \setlength{\parsep}{0pt}
  
}{\end{itemize}}

\newcommand{\gael}[1]{{\color{black} #1}}
\newcommand{\meinard}[1]{{\color{black} #1}}
\newcommand{\vincent}[1]{{\color{black} #1}}
\newcommand{\eric}[1]{{\color{black} #1}}

\title{\vspace{-6ex} \meinard{Model-Based} Deep Learning for Music Information Research}

\author{\emph{Ga\"el Richard$^{\star}$, Vincent Lostanlen$^{\dagger}$, Yi-Hsuan Yang$^{\ddagger}$, Meinard M{\"u}ller$^{\star\star}$}\vspace{-0.5ex} \\ \\ \small $^{\star}$LTCI, T{\'e}l{\'e}com Paris, Institut polytechnique de Paris, France, \url{gael.richard@telecom-paris.fr}\\ \small $^{\dagger}$Nantes Université, École Centrale Nantes, CNRS, Nantes, France, \url{vincent.lostanlen@ls2n.fr}\\ \small $^{\ddagger}$National Taiwan University, Taiwan, \url{yhyangtw@ntu.edu.tw}\\ \small $^{\star\star}$International Audio Laboratories Erlangen, Germany, \url{meinard.mueller@audiolabs-erlangen.de}\vspace{-2ex}}
\date{}

\begin{document}

%
%


\maketitle


\linespread{\maintextlinespacing}
\selectfont


\begin{abstract}
In this article, we investigate the notion of \textit{\meinard{model-based deep learning}} in the realm of music information research (MIR). \meinard{Loosely speaking, we refer to the term model-based deep learning for approaches that combine traditional knowledge-based methods with data-driven techniques, especially those based on deep learning, within a differentiable computing framework.}
%
%
In music, prior knowledge for instance related to sound production, music perception or music composition theory can be incorporated into the design of neural networks and associated loss functions. We outline three specific scenarios to illustrate the application of \gael{model-based deep learning} in MIR, demonstrating the implementation of such concepts and their potential.
\end{abstract}

\section*{Introduction}

Access to increasingly powerful supercomputing facilities, coupled with the availability of vast datasets has lead to a shift towards data-driven methodologies in nearly all areas of multimedia processing, including speech, music, and audio processing. The field has embraced end-to-end neural approaches, with a focus on directly addressing machine learning problems for raw acoustic signals (e.g., waveform-based signals captured directly from microphones). However, these approaches often overlook the specific nature and structure of the processed data. As a result, these complex models, some comprising up to hundreds of billions of parameters (such as recent large-language models), necessitate extensive training on large datasets to achieve effectiveness, albeit at the cost of limited interpretability and control.

\begin{figure}[t]
\centering
\includegraphics[width=14cm]{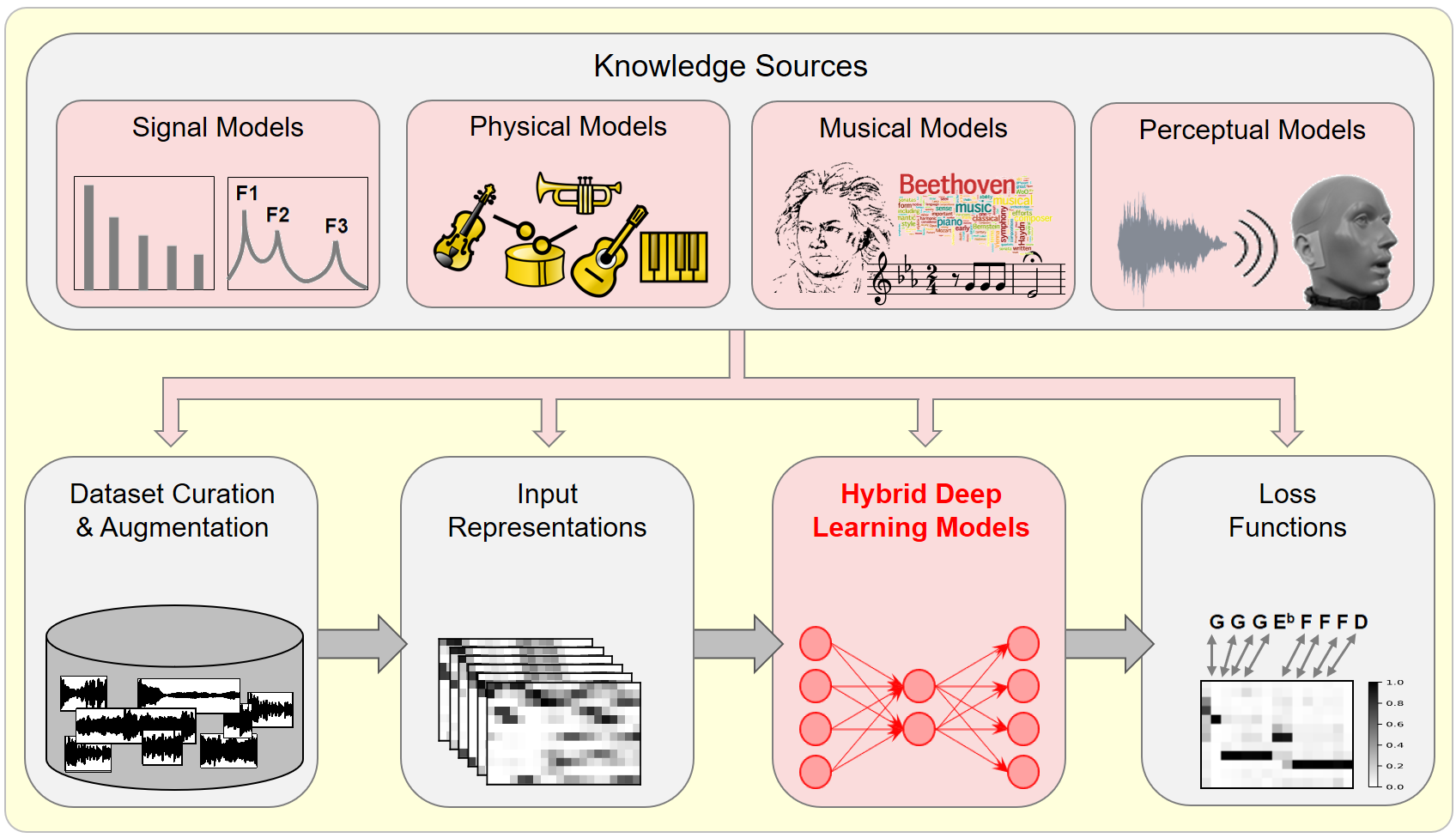}
\caption{Knowledge sources in MIR and their integration in the workflow of deep learning pipelines.}
\label{fig:teaser}
\end{figure}

This trend is also evident in Music Information Research (MIR), a field addressing applications such as music transcription, recognition, and generation, irrespective of whether the processed data consists of acoustic audio signals (e.g., music recordings) or symbolic music representations (e.g., encoded sheet music). In the realm of music, prior knowledge may relate to sound production (using an acoustic or physical model), music perception (based on a perceptual model), or music composition (using a musicological model). These models typically incorporate only a few parameters, facilitating the creation of systems that are both controllable and interpretable. 
Modern machine learning frameworks based on neural networks can leverage these models, \meinard{giving rise to umbrella terms such as \textit{model-based deep learning} sometimes also referred to \textit{hybrid deep learning}} (refer also to Figure~\ref{fig:teaser}).
%

%
While offering a precise definition of model-based (or hybrid) deep learning proves challenging,  \meinard{this term generally relates to approaches that combine traditional knowledge-based methods with data-driven deep-learning techniques.}
As detailed in~\cite{shlezinger2023model}, \meinard{such models} include concepts like model-based networks (where a knowledge model shapes the network architecture and aids in training) and DNN-based inference (where a neural network directly predicts the parameters of a knowledge model). \meinard{Taken literally, model-based (or hybrid) deep learning could encompass virtually all recent work beyond purely data-driven end-to-end systems or purely knowledge-based models.}
\meinard{In this paper, we employ the term \emph{model-based deep learning} when knowledge is integrated into the architecture, layers, and loss functions within a deep neural network (DNN) pipeline and within a differentiable computing framework.}
%
%
\meinard{In this context, the property of \emph{differentiability} is especially important as it enables the optimization of model parameters via gradient descent and its variants, see also the Box~\ref{box:Diff} for further explanations.}


The main objective of this article is to delve deeper into the concept of \meinard{model-based deep learning} within the MIR field and demonstrate its potential through three carefully selected illustrative application scenarios. This investigation encompasses scenarios where music is represented either as an audio signal or a symbolic sequence.

\section*{Knowledge Sources in MIR}

A widely accepted principle in MIR, as stated in~\cite{MuellerEKR11_MusicProcessing_IEEE-JSTSP}, is that \textit{``... to be successful, music audio signal processing techniques must be informed by a deep and thorough insight into the nature of music itself.''} We emphasize that \gael{model-based} deep learning holds significant potential, especially in the music domain, thanks to the richness and diversity of \textit{knowledge sources} for music signals (see Figure~\ref{fig:teaser}). These knowledge sources encompass signal models (e.g., encoding harmonic, melodic, or rhythmic properties), physical models (e.g., used for sound production of specific instruments), musical models (e.g., encoding specific music theory), and perceptual models (e.g., for mimicking the way humans perceive sounds). 

To contextualize \gael{model-based deep learning} in MIR, we discuss the main ways of integrating prior knowledge into deep learning pipelines (as illustrated in Figure~\ref{fig:teaser}). Firstly, musical knowledge is crucial for the careful assembly and annotation of datasets used in model training and testing. In this context, physical models can contribute to synthetic audio data generation and augmentation. Secondly, one can leverage knowledge about music signals or sound perception to derive informed feature representations as input to neural networks, such as mel spectrograms, harmonic constant-Q spectrograms~\cite{BittnerMSLB17_DeepSalience_ISMIR} or high-level audio features~\cite{Peeters2021}.
%
Furthermore, knowledge from traditionally engineered MIR pipelines can be utilized to customize the architecture of the neural network, as demonstrated in~\cite{DurandBDR17_DownbeatTrackEnsembleCNN_TASLP} for downbeat estimation or in~\cite{wu22ismir} to provide rhythm-related information for drum accompaniment generation. \eric{Lastly, musical or perceptual knowledge can guide the design of appropriate loss functions (e.g., the multiscale spectral loss~\cite{SchwaerM23_MultiScaleSpecLoss_IEEE-SPL}) or motivate the inclusion of additional loss terms~\cite{9052944}}.

\section*{Related Work on Hybrid Models for Music}

The domain of music has increasingly captured the attention of the signal processing community over the past two decades. For example, the special issue on \emph{Music Signal Processing}~\cite{MuellerEKR11_MusicProcessing_IEEE-JSTSP} reflects these activities during a phase when MIR began to gain recognition as an independent research area within the broader context of signal processing. As for more recent developments, the special issue on \emph{Recent Advances in Music Signal Processing}~\cite{MuellerPMV19_Editorial_IEEE-SPM} surveys novel developments across various areas of music processing with a focus on audio. While there are general review articles on hybrid signal processing techniques and model-based deep learning (see, e.g.,~\cite{Daw2017, shlezinger2023model}), only a few current works specifically address \gael{model-based} deep learning concepts within the context of music.

Several studies have introduced innovative neural network architectures that leverage a signal model, with one of the most noteworthy examples being the Differentiable Digital Signal Processing (DDSP) framework. Originally introduced in~\cite{EngelHGR20_DifferentiableDSP_ICLR}, the core concepts of the DDSP framework aim to seamlessly integrate traditional Digital Signal Processing (DSP) elements such as filters, oscillators, and reverberation into deep learning pipelines. This framework serves as a means to incorporate domain knowledge, represented by a well-defined signal model, into the neural network. As extensively discussed in the review in~\cite{hayes2023review}, DDSP-based methods offer several key advantages, including:

\begin{itemizePacked}
\item \textit{Data and computational efficiency:} The incorporation of the signal synthesis model introduces prior knowledge with only a few model parameters.
\item \textit{Interpretability:} The signal synthesis model is inherently interpretable.
\item \textit{Controllability:} The parameters of a sinusoidal--noise model are directly associated with sound components exhibiting distinct perceptual attributes such as pitch, loudness, and harmonicity, thereby serving as intuitive control parameters for synthesis.
\end{itemizePacked}


DDSP-based methods have the potential to produce high-quality audio synthesis results when the parametric signal model used is sufficiently expressive. Currently, DDSP-based models have been incorporated into various applications, including music and speech synthesis, audio effects generation, data augmentation, and source separation. Following~\cite{Schulze-Forster2023-TSALP}, we will further delve into the DDSP idea in our first application scenario below, describing a hybrid approach that integrates source--filter models for the separation of polyphonic choir recordings.
Before doing so, we would like to highlight several other noteworthy examples of \gael{model-based} deep learning approaches in the audio domain, without claiming to be exhaustive.

\begin{itemizePacked}
\item \textit{Coupling signal processing modules with deep learning:} This strategy is employed across various tasks, encompassing representation learning~\cite{Ravanelli2018}, audio synthesis~\cite{EngelHGR20_DifferentiableDSP_ICLR}, and phase recovery~\cite{Masuyama2021}. In the realm of music analysis, works such as~\cite{Pons2016,SchreiberM19_TempoKeyCNN_SMC} leverage domain knowledge to constrain the filter shapes of convolutional neural networks.

\item \textit{Designing neural network architectures rooted in physical models:} Illustrative examples of this approach include piano models~\cite{renault2022}, singing voice source--filter models~\cite{Schulze-Forster2023-TSALP}, and the use of differential equations for simulating sound effects~\cite{Wilczek2022}.

\item \textit{Imposing constraints on models with attention mechanisms:} This strategy is evident in diverse applications such as lyrics-driven audio source separation~\cite{Schulze2021-TSALP}, symbolic music completion~\cite{Guo2022}, and style transfer~\cite{Cifka2020}. In the latter, the attention mechanism plays a crucial role in facilitating the soft alignment of various knowledge sources, enabling the transfer of a given music piece's style to another musical excerpt.

\item \textit{Integrating perceptual side information for model conditioning:} This approach is exemplified in~\cite{Nistal2020}, where it is applied for drum synthesis.

\item  \textit{Obtaining DNN-based inference models for parameter estimation:} This method is employed in the context of handling complex non-differentiable black-box audio processing effects for music signal transformation, as discussed in~\cite{ramirez21icassp}.

\end{itemizePacked}

To further illustrate the potential of \gael{model-based} deep learning concepts for MIR applications, we will now explore three specific application scenarios of hybrid learning techniques inspired by music. Rather than addressing all possible aspects of integrating musical knowledge, our focus will be on musically informed network architectures and musically constrained loss functions.

\section*{Application: Music Source Separation Integrating a Source--Filter Model}
\label{sec:sourceseparation}

\begin{figure}[t]
\centering
\includegraphics[width=14cm]{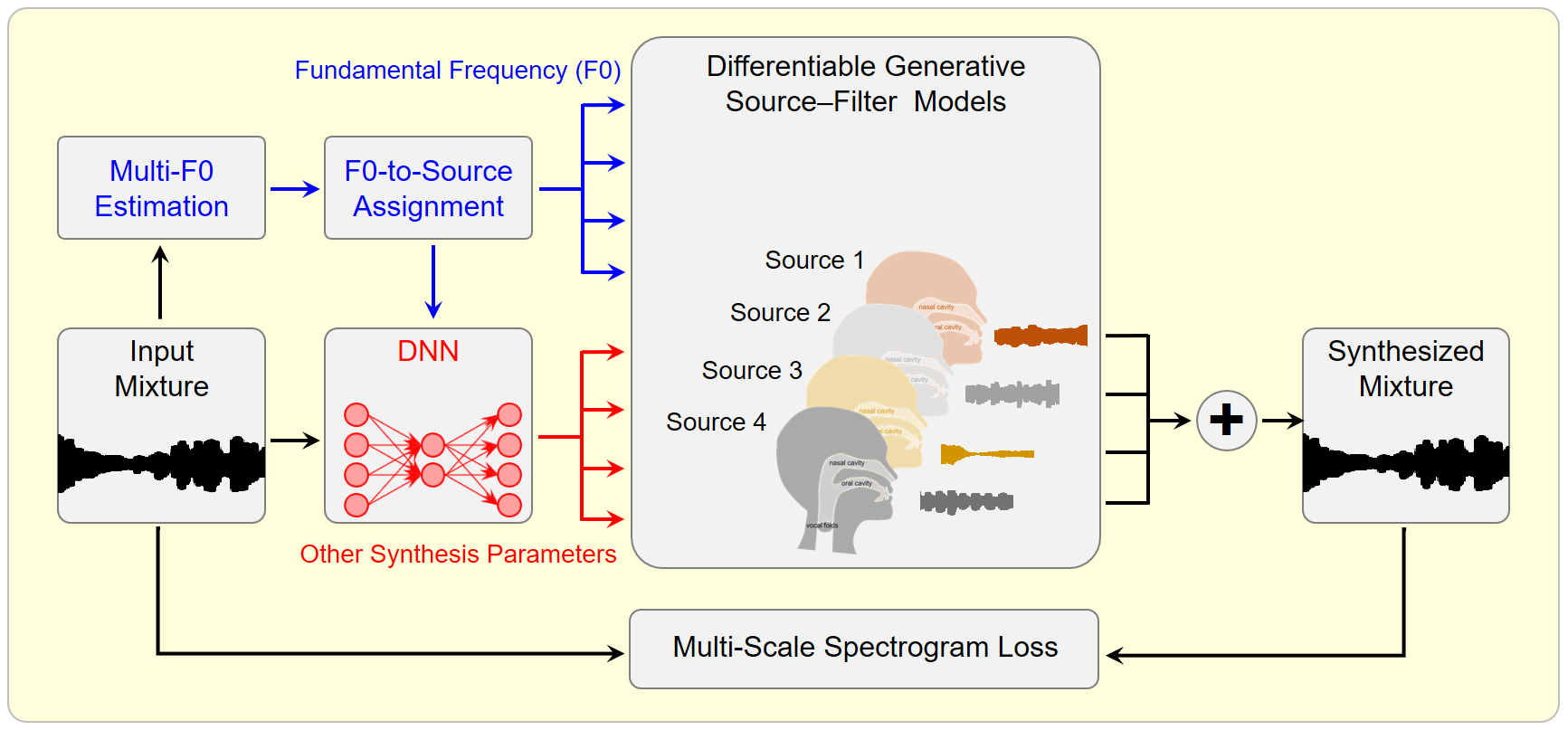}
\caption{{Unsupervised source separation method integrating source--filter models for singing voice production into a deep learning pipeline.}}
\label{Fig:unsupervised-SS}
\end{figure}

Our first application scenario serves as an illustrative example of a DDSP-like hybrid model designed for the unsupervised separation of vocals from polyphonic choir recordings. Following the approach proposed in~\cite{Schulze-Forster2023-TSALP}, the concept revolves around integrating differentiable parametric source--filter models, originally designed for speech production, into a data-driven DNN framework. In the musical choir context, each source is encoded through a source--filter model, playing the role of a singing voice production model. We will now delve into the details of this \gael{model-based} approach, following the illustration in Figure~\ref{Fig:unsupervised-SS}.

\begin{itemizePacked}
\item \textit{Task formulation:} 
The source separation problem addressed in this scenario deals with a polyphonic choir recording, a single-channel mixture containing the voices of multiple singers (for instance, four singers as shown in Figure~\ref{Fig:unsupervised-SS}). The unsupervised source separation method aims to untangle this complex mixture, separating it into individual singing voices or sources. In our example, the ultimate goal is to produce four distinct tracks, each corresponding to the voice of one singer, as if they had sung in isolation.

\item \textit{Source--filter singing voice production model:} 
Being the crucial component in facilitating the separation process, this component represents each source, corresponding to an individual singing voice, using a differentiable parametric model. The parameters of this model encapsulate crucial characteristics such as pitch, timbre, and loudness, offering the network a structured understanding of the nature of singing voices. 
Essentially, the source--filter model serves as a representation of how singing voices are produced: the vibration of the vocal folds generates an excitation signal that propagates in the nasal and oral cavities, acting as a resonating filter to shape the produced sound. The excitation signal is synthesized using a differentiable harmonics plus noise model, appropriate to represent one singer's voice characteristics. Short-term variations, e.g., due to articulations of words, are modeled by a parametric all-pole filter. 
During the separation process, the DNN utilizes the knowledge given by these differentiable source--filter models to guide the reconstruction of individual sources from the mixture. The source--filter models act as informed constraints, helping the neural network generate outputs that align with the expected characteristics of singing voices. 

\item \textit{Integration into deep learning pipeline:} 
The DNN is the overarching structure that learns to separate different sources from a polyphonic choir recording. During training, the DNN's objective is to reconstruct the observed mixture by estimating the parameters of each source, given their fundamental frequencies. In other words, the DNN learns a mapping that, given the fundamental frequencies of the sources, generates an output resembling the input mixture. It learns how each source contributes to the overall sound, considering the characteristics provided by the source--filter model.

\end{itemizePacked}

Let us assume a more conceptual viewpoint on the hybrid approach by introducing some mathematical notations. Let $\boldsymbol{x}$ denote the polyphonic music signal to be separated, and $K$ be the number of sources. For example, in Figure~\ref{Fig:unsupervised-SS} we have $K=4$ sources. 
For each source $k\in\{1,2,\ldots,K\}$, a non-learnable source--filter model $\boldsymbol{g}_k$ takes a parameter vector $\boldsymbol{\theta}_k$ as input and generates a synthesized source signal $\boldsymbol{x}_k=\boldsymbol{g}_k(\boldsymbol{\theta}_k)$. 
\gael{The non-learnable parameters $\boldsymbol{\theta}_k$ of a source $k$ include the fundamental frequency and the amplitudes of the harmonic plus noise components of the excitation signal, and the all-pole filter coefficients characterizing the resonating filter of the nasal and oral cavities.}
In our unsupervised scenario, we lack knowledge about the actual source signals, about the optimal parameter vectors, and the approximation quality of the synthesized source signals. The \gael{model-based} approach involves a DNN $\boldsymbol{f}_{\mathbf{W}}$ with learnable weights $\mathbf{W}$, aiming to estimate source--filter model parameters:
\begin{equation}
\boldsymbol{f}_{\mathbf{W}}(\boldsymbol{x})=\boldsymbol{\tilde{\theta}}=(\boldsymbol{\tilde{\theta}}_1,\ldots,\boldsymbol{\tilde{\theta}}_K),
\end{equation}
such that the superposition $\boldsymbol{\tilde{x}}=\sum_{k=1}^K \boldsymbol{\tilde{x}}_k$  of the estimated synthesized source signals $\boldsymbol{\tilde{x}}_k=\boldsymbol{g}_k(\boldsymbol{\tilde{\theta}}_k)$ closely approximates the original signal $\boldsymbol{x}$. In essence, the success of the hybrid model is assessed by how well the synthesized mixture signal $\boldsymbol{\tilde{x}}$ approximates the original mixture signal $\boldsymbol{x}$. 

To optimize the network parameters $\mathbf{W}$ of the DNN $\boldsymbol{f}_{\mathbf{W}}$, a loss function is essential to quantify the dissimilarity between the original signal $\boldsymbol{x}$ and the estimated signal $\boldsymbol{\tilde{x}}$. Instead of employing a waveform-based loss function, we adopt a multiscale spectral loss, as used in~\cite{EngelHGR20_DifferentiableDSP_ICLR}. This loss function, by assessing time--frequency representations across multiple frequency scales, demonstrates sensitivity to both global and local spectral characteristics that correlate with the perception of sound similarity. Further details on this sensitivity are discussed in the subsequent application scenario.
To formulate this loss, let $\boldsymbol{\Phi}_N$ denote an operator that transforms a signal $\boldsymbol{x}$ into a magnitude spectrogram  $\boldsymbol{\Phi}_N(\boldsymbol{x})$. Here, $N$ is a parameter determining the operator's characteristics, such as the window size used in the spectrogram computation. Let $\mathcal{N}=\{2048, 1024, 512, 256, 128, 64\}$ represent a set of possible window sizes. With this notation, the multiscale spectral loss is defined as follows:
\begin{equation}
\label{eq:MSS}
     \mathcal{L}_{\boldsymbol{\theta}}(\mathbf{W})   =  
       \sum_{N\in\mathcal{N}} \hspace{0.1cm} ( \Vert \boldsymbol{\Phi}_N(\boldsymbol{x}) - \boldsymbol{\Phi}_N(\boldsymbol{\tilde{x}}) \Vert_1 + \Vert \log(\boldsymbol{\Phi}_N(\boldsymbol{x})) - \log(\boldsymbol{\Phi}_N(\boldsymbol{\tilde{x}})) \Vert_1 ).
 \end{equation}
For additional insights and a critical discussion of this loss function, please refer to~\cite{SchwaerM23_MultiScaleSpecLoss_IEEE-SPL}.

To summarize, the neural network is trained to output parameters for source--filter models for each of the sources to be separated. In the loss function, the comparison is solely between the superposition $\boldsymbol{\tilde{x}}$ of the individually synthesized sources and the original signal $\boldsymbol{x}$. In essence, the individual sources are synthesized through the source--filter production models, enabling the learning of the complete source separator directly from the mixture signals, without explicit knowledge of the individual sources. This results in a fully unsupervised \gael{model-based} deep learning approach for audio source separation.

Moreover, the source--filter production model, expressed with a few meaningful parameters $\boldsymbol{\tilde{\theta}}$, allows for direct source transformations related to excitation (e.g., pitch changes, voice breathiness) or the filter (e.g., timbre change, phoneme alteration). In particular, it facilitates the separation of homogeneous sources, such as those with similar timbre, as found in choir recordings.
Finally, this approach leads to substantial efficiency improvements in terms of required training data and model simplicity, \meinard{as elaborated in \cite{Schulze-Forster2023-TSALP} using multiple objective metrics}.
However, \gael{model-based} deep learning approaches that rely on prior knowledge of source production models also encounter significant challenges. The current model, tailored for singing voices, does not readily generalize to other source types.
\meinard{To enable general music source separation, including instruments with percussive and transient components (such as drums) or instruments capable of producing multiple voices (e.g., piano), specific source production models will be needed (see~\cite{renault2022} for instance).}
%
%
Additionally, some parameters, such as the number of sources and fundamental frequencies of each source over time, are estimated externally. For an approach addressing some of these challenges, we refer to~\cite{Richard2024-Icassp}.
\meinard{Links to sounds examples illustrating the performances of some of these methods can be found in~\cite{Schulze-Forster2023-TSALP,Richard2024-Icassp}}.


\section*{Application: Perceptual Sound Matching \vincent{with Joint Time--Frequency Scattering}}


As previously mentioned, comparing two sound signals, denoted as $\boldsymbol{x}$ and $\boldsymbol{\tilde{x}}$, solely based on their waveforms can pose challenges attributed to factors such as susceptibility to time shifts, spectral deviations, and phase sensitivity. In other words, while the waveform distance $\Vert \boldsymbol{\tilde{x}} - \boldsymbol{x} \Vert$ may be large, the two signals may be very similar from a human perception point of view. To address these issues, one typically converts the signals to be compared into time--frequency representations (TFRs) and then compares the signals based on these representations using suitable distance measures. In the preceding discussion, we have already seen an example using magnitude spectrograms as TFRs and the multiscale spectral loss as a distance function (see Equation~\ref{eq:MSS}).
More generally, let $\boldsymbol{\Phi}$ denote an operator that transforms a signal $\boldsymbol{x}$ into a TFR, expressed as $\boldsymbol{\Phi}(\boldsymbol{x})$. The desired property of $\boldsymbol{\Phi}$ is that the Euclidean (or another suitable) distance $\Vert\boldsymbol{\Phi}(\boldsymbol{\tilde{x}}) - \boldsymbol{\Phi}(\boldsymbol{x})\Vert$ should be small precisely when the two sound signals $\boldsymbol{x}$ and $\boldsymbol{\tilde{x}}$ are perceived similarly by a human listener. Although the magnitude spectrogram operator handles phase, translation, and slight pitch changes effectively, its alignment with human perception is only partial. 

We now delve deeper into these aspects by considering an application scenario referred to as perceptual sound matching. Following~\cite{han2024learning}, we examine a scenario where the objective is to match a given sound $\boldsymbol{x}$ with a synthesized version $\boldsymbol{\tilde{x}}$ in a perceptually convincing fashion. 
More precisely, using similar notation as introduced earlier, let $\boldsymbol{g}$ be a synthesizer model that takes a parameter vector $\boldsymbol{\theta}$ to generate a synthesized sound $\boldsymbol{g}(\boldsymbol{\theta})$. 
Given an arbitrary sound $\boldsymbol{x}$, the sound matching task aims to find a parameter vector $\boldsymbol{\tilde{\theta}}$ such that the synthesized sound $\boldsymbol{\tilde{x}}=\boldsymbol{g}(\boldsymbol{\tilde{\theta}})$ is judged by a human listener to resemble  $\boldsymbol{x}$. 
As discussed in~\cite{han2024learning}, sound matching has applications in automatic music transcription, virtual reality, and audio engineering. A particularly interesting scenario is when $\boldsymbol{g}(\boldsymbol{\theta})$ solves a known partial differential equation (PDE) with respect to the variables $\boldsymbol{\theta}$. In this case, the parameter vector $\boldsymbol{\theta}$ unveils critical design choices in acoustical manufacturing, including the shape and material properties of the resonator.

In the following, we assume that the original signal $\boldsymbol{x}$ is also generated by the synthesizer model, i.e., $\boldsymbol{x} = \boldsymbol{g}(\boldsymbol{\theta})$ for some parameter vector $\boldsymbol{\theta}$. This parameter serves as the target to be estimated by a DNN $\boldsymbol{f}_{\mathbf{W}}$ with weights $\mathbf{W}$. In other words, our objective is to train the model to produce an estimate $\boldsymbol{\tilde{\theta}} = \boldsymbol{f}_{\mathbf{W}}(\boldsymbol{x})$ for $\boldsymbol{\theta}$. Instead of defining a loss in the parameter space, such as $\Vert \boldsymbol{\theta} - \boldsymbol{\tilde{\theta}}\Vert^2$, the study~\cite{han2024learning} demonstrates the benefits of formulating the loss function in a suitably defined representation space of the sound signals:
\begin{equation}
\label{eq:loss}
\mathcal{L}_{\boldsymbol{\theta}}(\mathbf{W}) = 
 \Vert \boldsymbol{\Phi}(\boldsymbol{x}) - \boldsymbol{\Phi}(\boldsymbol{\tilde{x}}) \Vert^2 
=
\Vert (\boldsymbol{\Phi}\circ\boldsymbol{g})(\boldsymbol{\theta}) - (\boldsymbol{\Phi}\circ\boldsymbol{g}\circ\boldsymbol{f}_{\mathbf{W}}\circ\boldsymbol{g})(\boldsymbol{\theta}) \Vert^2.
\end{equation}
%
%
\vincent{For this purpose, the time--frequency representation $\boldsymbol{\Phi}$ needs to satisfy two properties.
First, it should closely correlate with the human perception of sound, taking into account dependencies on the specific signal type and requirements of the considered use case.
Second, $\boldsymbol{\Phi}$ should be implemented in a programming framework that offers automatic differentiation, thus allowing the computation of the gradient of $\mathcal{L}_{\boldsymbol{\theta}}$ with respect to neural network weights $\boldsymbol{\mathbf{W}}$.}

%
%
\meinard{To mitigate some of the shortcomings of the multiscale spectral loss (see Equation~\ref{eq:MSS}), one may define $\boldsymbol{\Phi}$ using the concept of joint time--frequency scattering (JTFS)}, a wavelet-based operator which extracts spectro--temporal modulations at various scales and rates~\cite{lostanlen2021time}.
JTFS is grounded in a simplified mathematical model of spectro--temporal receptive fields, which constitute the early stages of auditory processing in the mammalian cortex. Behavioral experiments involving classically trained composers have validated that the Euclidean distance between JTFS coefficients accurately predicts the perceived timbre dissimilarity between musical notes, including extended playing techniques~\cite{lostanlen2021time}.
These insights prompted~\cite{vahidi2023mesostructures} to implement JTFS as a differentiable operator for hybrid deep learning and distribute it through the Kymatio open-source software (\url{https://github.com/kymatio/kymatio}).

However, JTFS has a drawback due to its higher computational cost compared to conventional spectrograms. Even with GPU acceleration, the majority of the time during neural network training with JTFS-based DDSP is spent on backpropagation through $\boldsymbol{\Phi}$ rather than updating the weights $\boldsymbol{\mathbf{W}}$.
Importantly, we can take advantage of the hybrid structure of the loss function $\mathcal{L}_{\boldsymbol{\theta}}$ to accelerate the computation of its gradient. Intuitively, we employ a surrogate dissimilarity metric that approximates the JTFS distance within a neighborhood of each ground truth sample $\boldsymbol{\theta}$.
As demonstrated in~\cite{han2024learning}, this approximation can be derived by utilizing the first-order term of the Taylor expansion of $(\boldsymbol{\Phi}\circ\boldsymbol{g})(\boldsymbol{\theta})$ around the training sample $\boldsymbol{\theta}$. 
%
%
\vincent{This derivation results in a quadratic approximation of the loss function $\mathcal{L}_{\boldsymbol{\theta}}$ (see Equation~\ref{eq:loss}), in~\cite{han2024learning} referred to as the \emph{Perceptual--Neural--Physical} (PNP) loss function:
\begin{equation}
\mathcal{L}_{\boldsymbol{\theta}}^{\mathrm{PNP}}(\mathbf{W}) =
\big(f_{\boldsymbol{W}}(\boldsymbol{x}) -\boldsymbol{\theta}\big)^{\top}
\cdot
\mathbf{M}(\boldsymbol{\theta})
\cdot
\big(f_{\boldsymbol{W}}(\boldsymbol{x})-\boldsymbol{\theta}\big),
\end{equation}
where $\mathbf{M}(\boldsymbol{\theta}) =
\boldsymbol{\nabla}_{\boldsymbol{\Phi}\circ\boldsymbol{g}}(\boldsymbol{\theta})^{\top}
\cdot \boldsymbol{\nabla}_{\boldsymbol{\Phi}\circ\boldsymbol{g}}(\boldsymbol{\theta})$ is a small matrix that may be precomputed independently of $\mathbf{W}$}.
Experiments in this study revealed that the linearization procedure provided by the PNP loss achieves a 100-fold speedup during gradient descent compared to the JTFS loss. Moreover, numerical simulations on a dataset of percussive sounds demonstrated that PNP achieves state-of-the-art results in perceptual sound matching at scale, measured in terms of JTFS Euclidean distance, outperforming both parameter loss and spectrogram-based loss functions.
\meinard{For sound examples, we refer to \url{https://lylyhan.github.io/perceptual_neural_physical}.}
\section*{Application: Symbolic-Domain Music Generation Incorporating Musicological Knowledge}

In our third application scenario, we consider automatic music generation which aims to autonomously create novel pieces of music, be it in the symbolic domain (using formats like MIDI) or the audio domain (as acoustic signals). Tasks within this realm encompass harmonizing melodies, generating music arrangements, creating accompaniments, and converting text to music. As these tasks are 
about writing music, they can benefit from human knowledge 
of
how music is composed, arranged, performed, and produced. Consequently, \meinard{model-based and hybrid deep learning approaches}  have emerged, exploiting musicological models to enhance various stages in the workflow of deep learning pipelines, as illustrated in Figure~\ref{fig:teaser}  
and exemplified below.

\meinard{
For example, musicological models can enhance the input \emph{data representation}. While conventional formats like MIDI encode basic musical parameters such as notes, durations, and dynamics, musicological models capture higher-level structures like motifs, themes, harmonic progressions, and rhythmic patterns~\cite{huang20remitransformer}, providing the composition model with a richer and more contextually informed input.
Similarly, musicological models contribute to \emph{structured learning} by understanding the interdependencies among musical events within a composition. This allows the model to generate music that reflects broader musical contexts and relationships, such as harmonic connections, thematic variations, and recurring patterns~\cite{janssen13CMMRa}, resulting in a more coherent and structured composition.
Finally, \emph{refining loss functions} with musicological models that incorporate principles from music theory~\cite{jaques17iclrworkshop} can enhance the optimization process of the generation model. For example, the loss function could prioritize thematic coherence or penalize deviations from established harmonic structures.
}

\begin{figure}[t]
\centering
\includegraphics[width=11cm]{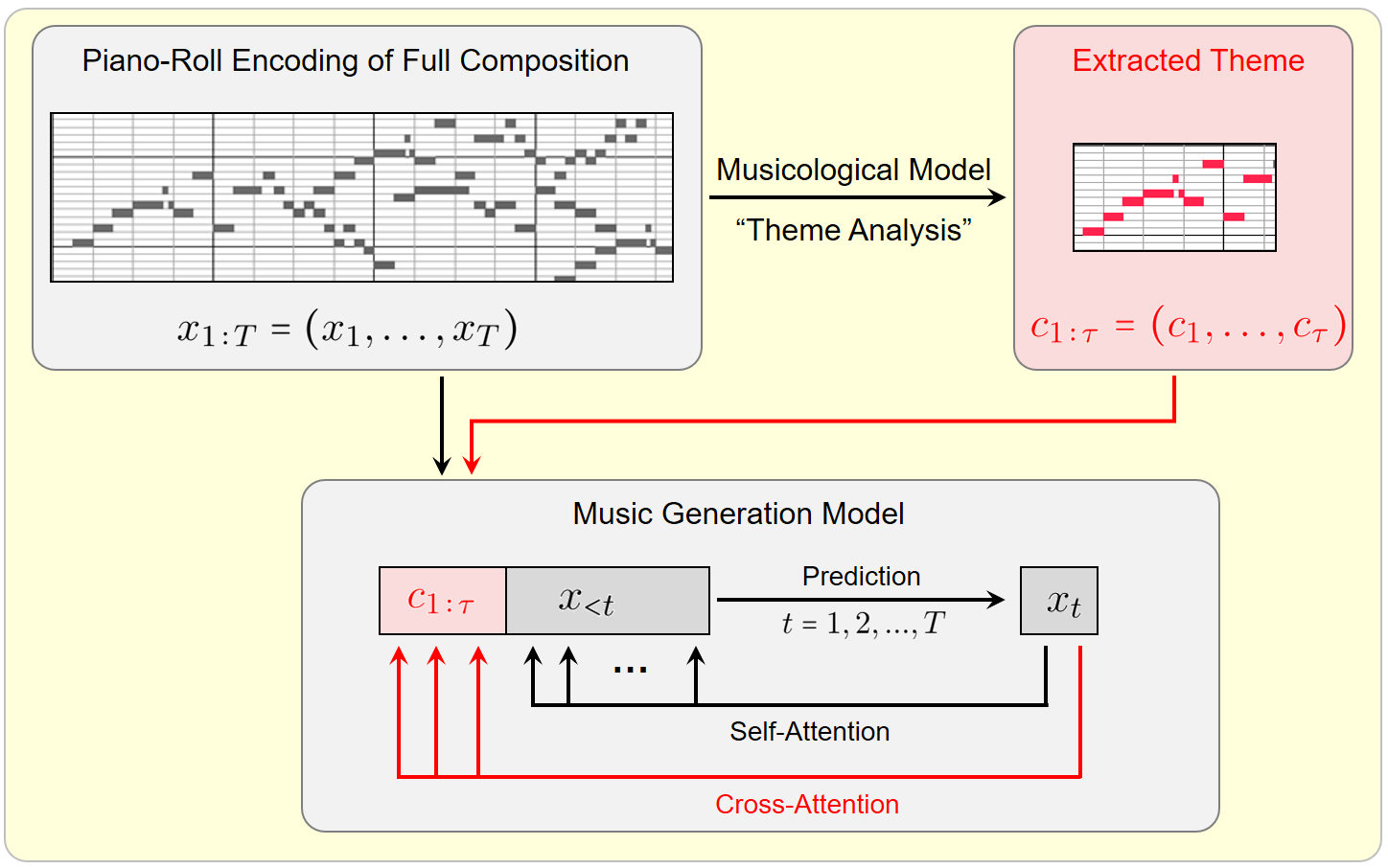}
\caption{Training an autoregressive music generation approach conditioned on a musical theme. The theme may be specified by a user or extracted from a given composition using a musicological model.
}
\label{Fig:MusicGeneration}
\end{figure}

\meinard{In the following, we take a close look at the Theme Transformer model~\cite{shih22tmm}, which serves as an example of how some of these concepts can be practically applied. Generally, Transformers learn dependencies among elements within a sequence by addressing the task of predicting the next element. In particular, given a sequence $x_{<t}=(x_1,\ldots,x_{t-1})$, the following conditional probability function is learned:}
\begin{equation}
    p(x_t \vert x_{<t}) \,.
    \label{eq:uncond}
\end{equation}
This general methodology 
can 
be applied to tokenized sequences of MIDI music.
For example, a piano composition provided in the form of a single-track MIDI file (see Figure~\ref{Fig:MusicGeneration}), can be converted into a token sequence. This sequence comprises 
tokens that convey information about the pitch, onset time, duration, and dynamic of each MIDI note event~\cite{huang20remitransformer}. 

However, empirical findings suggest that Transformer models solely focused on optimizing the criterion expressed by Equation~\ref{eq:uncond} cannot guarantee the generation of sequences with a repetitive structure as seen in real-world music~\cite{shih22tmm}.
While a generated piece may exhibit harmonic and stylistic coherence over a short time span, it lacks the development of a central `musical idea' over the course of the entire piece. Without additional incentives, the music might drift into randomness, deviating from its initial phrases without a clear direction as the sequence lengthens. This is undesirable since, unlike in text or other sequential data, repetition is a fundamental aspect of music.

In the realm of music, themes (or typically shorter motifs) are distinctive melodic or rhythmic patterns that encapsulate essential musical ideas. These recurring elements play a crucial role in defining the fundamental identity of a musical composition. Importantly, themes are not simply repeated but reoccur in musically modified forms, accentuating their musical significance \cite{bribitzer-stull_2015}.
The Theme Transformer model, proposed in~\cite{shih22tmm}, incorporates a musicological model that identifies the central musical theme of the piece under consideration. When provided with a piece represented as a token sequence $x_{1:T}$, the musicological model $\mathcal{M}$ has the task of uncovering the intrinsic musical theme $\mathcal{M}(x_{1:T})$. This theme is then represented as another token sequence $c_{1:\tau}=(c_1,\dots,c_\tau)$, where $\tau \ll T$.
The Theme Transformer dataset includes piano covers of polyphonic pop music, featuring arrangements for both the melody line and its accompanying part played on the piano. In this dataset, the full composition is given as a single-track MIDI file represented by the sequence $x_{1:T}$. Furthermore, the identified theme $c_{1:\tau}$ is supposed to be monophonic, being part of the melody line and forming a subset of $x_{1:T}$, as illustrated in Figure~\ref{Fig:MusicGeneration}.
Moreover, while the theme usually repeats multiple times in a musical piece, $c_{1:\tau}$ encodes a single occurrence of the theme considered to be representative.

\begin{figure}[t]
\centering
\includegraphics[width=12cm]{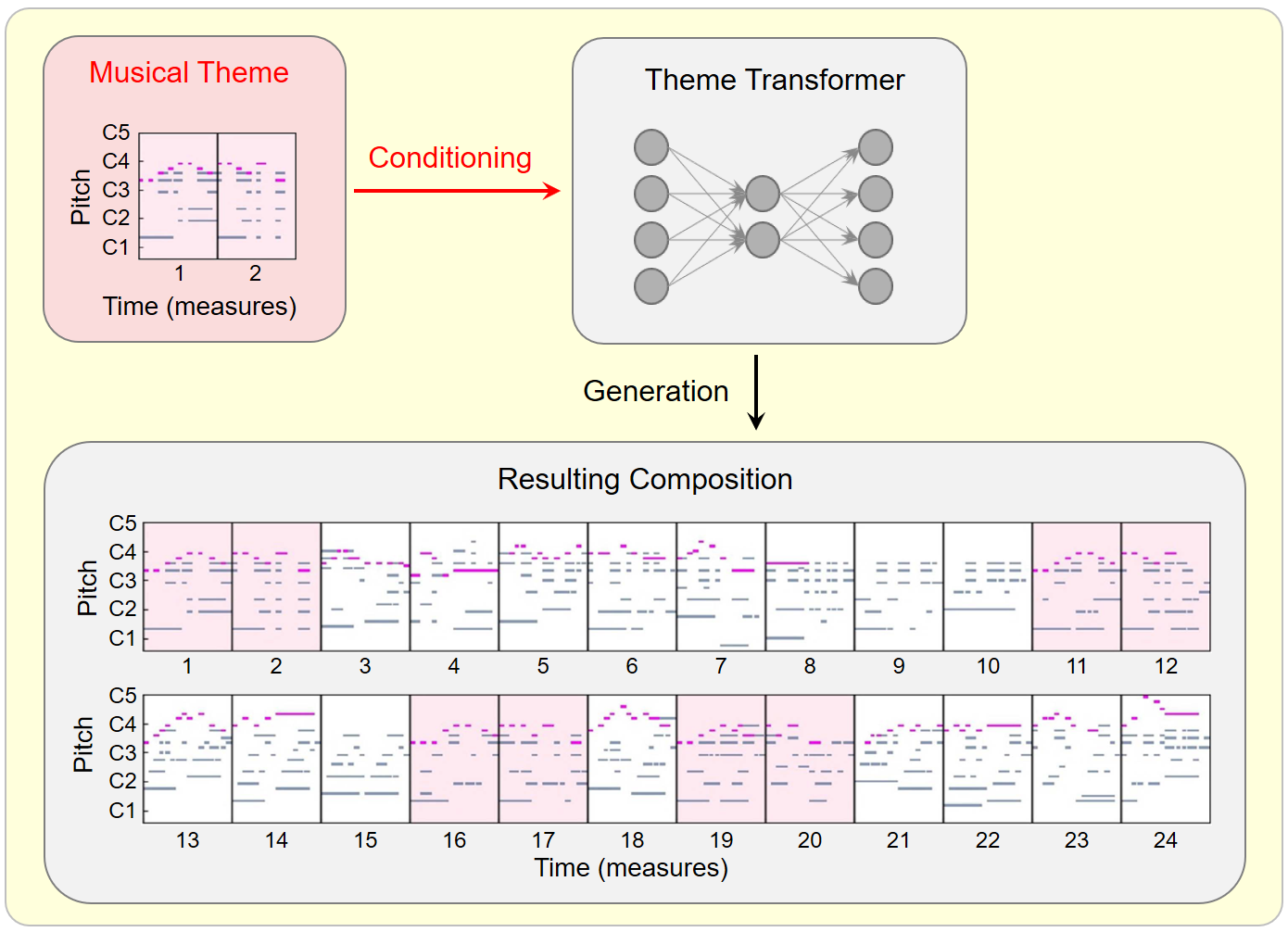}
\caption{Piano roll representation of the first 24 measures of a composition generated by the Theme Transformer, conditioned on a specified theme. The shaded pink regions represent the theme, the melody is highlighted in magenta, and the accompaniment is depicted in grey.
}
\label{Fig:MusicGeneration_example}
\end{figure}

Given 
knowledge of musical themes within the training set, the Theme Transformer models the following conditional probability function, taking both $x_{<t}$ and $c_{1:\tau}$ as conditions:
%
\begin{equation}
    p(x_t \vert x_{<t}, c_{1:\tau}) \,.
    \label{eq:cond}
\end{equation}
Specifically, as illustrated in Figure~\ref{Fig:MusicGeneration}, the Theme Transformer employs two 
memory networks: one self-attending to $x_{<t}$ and the other cross-attending to $c_{1:\tau}$. To balance information from these two memory networks, the model incorporates delicate designs, including a gating mechanism, theme-related tokens, and theme-aware positioning encoding \cite{shih22tmm}. 
At inference time of generating a new music piece, users can provide the target musical theme $c_{1:\tau}$ instead of extracting it from an existing piece. 

Comparative objective and subjective evaluations have shown that the Theme Transformer trained with the musically-motivated formulation (Equation~\ref{eq:cond}) outperforms the conventional approach\eric{\cite{huang20remitransformer}} (Equation~\ref{eq:uncond}), resulting in music with a more pronounced repetitive structure and enhanced overall musical quality~\cite{shih22tmm}. 
\meinard{For example, in a user study involving 17 listeners, samples generated by the conventional approach scored an average of 2.39 for `theme repetitiveness' and 2.78 for `overall musical quality' on a 5-point Likert scale (from 1 to 5, with higher being better). In comparison, the Theme Transformer scored 3.39 and 3.16, respectively, both of which are closer to the scores received by the original musical pieces (3.37 and 3.41).}
%

As an example, Figure~\ref{Fig:MusicGeneration_example} shows the MIDI file in the form of a piano roll representation, depicting the first 24 measures of a composition generated by the Theme Transformer, alongside the musical theme used for conditioning the generation process. In this snippet, the theme appears four times, each with variations in both the melody line and accompaniment. 
\eric{More examples can be found at \url{https://atosystem.github.io/ThemeTransformer/}.}

%
%
\meinard{While the Theme Transformer focuses on theme consistency, one research path is to integrate musicological models to improve other aspects of the generated music, such as theme variation and development.}

\section*{Conclusions and Future Work}


In this article, we have delved into the concept of \gael{model-based} deep learning, emphasizing its significance in the realm of music information research (MIR). By seamlessly combining knowledge-based and data-driven strategies within a differentiable deep neural framework, \gael{model-based} models leverage the diverse knowledge sources available for music signals. 
As discussed in the three application scenarios, the integration of model-based knowledge offers distinct advantages over purely data-driven paradigms. Firstly, we showed how the integration  of differentiable source--filter models facilitate the untangling of polyphonic choir recordings, offering a fully unsupervised approach that leverages knowledge of sound production. Secondly, we investigated how differentiable time–frequency representations could lead to perceptually motivated loss functions in the context of sound matching. Thirdly, we explored how to enhance a symbolic music generation system by incorporating higher-level musical structures through the integration of musicological models via conditioning.
We concluded that the incorporation of knowledge-based models characterized by a few interpretable parameters enhances explanability, controllability, and resource efficiency. In particular, the utilization of prior knowledge mitigates the reliance on extensive datasets during neural model training, facilitating the creation of smaller, more compact neural architectures without compromising effectiveness. 

Nonetheless, combining knowledge-based and data-based strategies poses considerable challenges. Firstly, these strategies may inherently contradict each other, necessitating specialized and well-crafted learning paradigms for finding an optimal compromise between the two strategies. Utilizing attention mechanisms emerges as a promising approach in this context, offering a versatile means to seamlessly combine information from both knowledge-based and data-based sources.
Another challenge lies in the fact that conventional sources of knowledge may be too specific, hindering generalization to novel situations. In addressing this, future work should concentrate on designing flexible and expressive knowledge-based models that can be smoothly integrated as differentiable modules into neural architectures. In essence, knowledge sources should guide the learning process without imposing overly strict constraints, ensuring the retention of the power of data-driven learning.

\section*{Acknowledgments}

Ga\"el Richard was supported by the European Union (ERC, HI-Audio, 101052978). Views and opinions expressed are however those of the author(s) only and do not necessarily reflect those of the European Union or the European Research Council. Neither the European Union nor the granting authority can be held responsible for them.
%
%
Yi-Hsuan Yang was supported by the National Science and Technology Council of Taiwan (NSTC 112-2222-E-002-005-MY2).
Meinard M{\"u}ller was funded by the Deutsche Forschungsgemeinschaft (DFG, German Research Foundation) under Grant No. 500643750 (MU 2686/15-1). The International Audio Laboratories Erlangen are a joint institution of the Friedrich-Alexander-Universit{\"a}t Erlangen-N{\"u}rnberg (FAU) and the Fraunhofer-Institut f{\"u}r Integrierte Schaltungen IIS.

\section*{Authors}
\label{sec:authors}

\noindent {\bf Ga\"el Richard } (gael.richard@telecom-paris.fr) 
received the State Engineering degree from Telecom Paris, France in 1990, the Ph.D. degree and Habilitation from University of Paris-Saclay respectively in 1994 and 2001. After the Ph.D. degree, he spent two years at Rutgers University, Piscataway, NJ, in the Speech Processing Group of Prof. J. Flanagan. From 1997 to 2001, he successively worked for Matra, Bois d’Arcy, France, and for Philips, Montrouge, France. He then joined Telecom Paris, where he is now a Full Professor in audio signal processing. He is also the co-scientific director of the \text{Hi! PARIS} interdisciplinary center on AI and Data analytics. He is a coauthor of over 250 papers and inventor in 10 patents. His research interests are mainly in the field of speech and audio signal processing and include topics such as source separation, machine learning methods for audio/music signals and music information retrieval. He is a fellow member of the IEEE, and was the chair of the IEEE SPS TC for Audio and Acoustic Signal Processing (2021-2022). He received, in 2020, the Grand prize of IMT-National academy of science.  In 2022, he was awarded of an advanced ERC grant of the European Union for a project on machine listening and artificial intelligence for sound.

\medskip

\noindent {\bf Vincent Lostanlen} (vincent.lostanlen@cnrs.fr) received the Ph.D. degree in Computer Science from École normale supérieure in 2017. Since 2020, he is a scientist at Centre national de la recherche scientifique (CNRS), affiliated to the Laboratoire des sciences du numérique de Nantes (LS2N).
Between 2017 and 2020, he was a postdoctoral researcher at the Cornell Lab of Ornithology and a visiting scholar at New York University (NYU).
His research interests include machine learning for music signal processing.
He is a core maintainer of two open-source software projects in audio signal processing: Librosa and Kymatio.

\medskip

\noindent {\bf Yi-Hsuan Yang} (yhyangtw@ntu.edu.tw) received the Ph.D. degree in Communication Engineering from National Taiwan University in 2010. Since February 2023, he has been with the College of Electrical Engineering and Computer Science, National Taiwan University, where he is a Full Professor. 
His research interests include automatic music generation, music information retrieval, artificial intelligence, and machine learning. 
He was a recipient of the 2011 IEEE Signal Processing Society Young Author Best Paper Award, 
and the 2019 Multimedia Rising Stars Award from the IEEE International Conference on Multimedia Expo. 
He was 
an Associate Editor for the IEEE Transactions on Multimedia and IEEE Transactions on Affective Computing, both from 2016 to 2019, and an author of the book ``Music Emotion Recognition'' (CRC Press 2011). 
Dr. Yang is a senior member of the IEEE.

\medskip

\noindent {\bf Meinard M\"uller} (meinard.mueller@audiolabs-erlangen.de) received the Diploma degree (1997) in mathematics and the Ph.D. degree (2001) in computer science from the University of Bonn, Germany. 
Since 2012, he has held a professorship for Semantic Audio Signal Processing at the International Audio Laboratories Erlangen, a joint institute of the Friedrich-Alexander-Universität Erlangen-Nürnberg (FAU) and the Fraunhofer Institute for Integrated Circuits IIS. His recent research interests include music processing, music information retrieval, audio signal processing, and motion processing. He was a member of the IEEE Audio and Acoustic Signal Processing Technical Committee (2010-2015), a member of the Senior Editorial Board of the IEEE Signal Processing Magazine (2018-2022), and a member of the Board of Directors, International Society for Music Information Retrieval (2009-2021, being its  president in 2020/2021). In 2020, he was elevated to IEEE Fellow for contributions to music signal processing. 
Furthermore, he wrote a textbook titled ``Fundamentals of Music Processing'' (Springer-Verlag, 2021), along with a comprehensive collection of educational Python notebooks designed for teaching and learning audio signal processing using music as an instructive application domain (\url{https://www.audiolabs-erlangen.de/FMP}).
%
%

\linespread{\referenceslinespacing}
\selectfont

{
\small
\bibliographystyle{IEEEtran}
\bibliography{referencesMeinard,referencesMeinard_New,referencesGael,referencesVincent,referencesEric}
}

\appendix

\linespread{\appendixlinespacing}  
\selectfont
\normalsize

\section{Importance of Differentiability}
\label{box:Diff}

\meinard{
Differentiability is essential in deep learning as it enables the optimization of model parameters via gradient descent and backpropagation. For gradient calculation, all components of a deep learning pipeline, including the loss function, activation functions, and network layers, must be differentiable. In model-based or hybrid pipelines that blend deep neural networks with traditional physical models and digital signal processing components, maintaining differentiability ensures the training process can make small, continuous adjustments to model parameters and weights, thereby minimizing the loss function. The backpropagation algorithm efficiently computes these gradients, propagating error gradients through the network and integrated models. Powerful tools and frameworks like PyTorch and TensorFlow leverage differentiability for automated gradient computation, streamlining the training of complex models.
}

\end{document}